\documentclass[10pt,preprint,showpacs]{revtex4}
\usepackage{amsmath}
\usepackage{graphicx}
\usepackage{amssymb, graphics}
\usepackage{amsfonts}

\newcommand{\rd}{\mathrm{d}}

\begin{document}
\title{Scalar perturbation of the viscosity dark fluid cosmological model}
\author{Xu Dou$^{1}$}
\email{dowxdou@gmail.com}
\author{Xin-He Meng$^{1,2}$}
\email{xhm@nankai.edu.cn}

\affiliation{$^{1}$Department of physics, Nankai University, Tianjin
300071, China\\$^{2}$Kavli Institute of Theoretical Physics
China,\\Chinese Academy of Sciences, Beijing 100190, China}

\begin{abstract}
A general equation of state is used to model unified dark matter and
dark energy (dark fluid), and it has been proved that this model is
equivalent to a single fluid with time-dependent bulk viscosity. In
this paper, we investigate scalar perturbation of this viscosity
dark fluid model. For particular parameter selection, we find that
perturbation quantity can be obtained exactly in the future
universe. We numerically solve the perturbation evolution equations,
and compare the results with those of $\Lambda$CDM model.
Gravitational potential and the density perturbation of the model
studied here have the similar behavior with the standard model,
though there exists significant value differences in the late
universe.

\end{abstract}


\maketitle

\section{Introduction}

Astrophysics and cosmology observations in recent years delineate
the cosmological picture on its constituents more and more
accurately, i.e, the precision cosmology era comes. Except for the
long standing puzzling dark matter component, an unknown cosmic
matter-energy constituent refereed to the so called dark energy may
also exist that accelerates our universe expansion now, which
contradicts with our traditional attitude on the behavior of
conventional matter but it is eventually confirmed by recent
observations like SNe Ia \cite{sn1} \cite{sn2} and CMB observations
\cite{CMB}. The cosmological dark sector, often divided as the
mysterious DM and DE sectors respectively, takes around $95\%$ of
total energy budget of our universe. The concord $\Lambda$CDM model
could be consistent with most of global astrophysics observational
results. But the introduction of the cosmological constant
simultaneously results in the yet to answer problems related
directly to how to understand the fundamental physics theory, like
the fine-tuning and the coincidence problems respectively. At the
same time, ``most'' of course does not equal to ``all'', some
astrophysics problems still need be clarified and solved in the
framework of the $\Lambda$CDM model \cite{six}, such as the the core
singularities of the cold matter halo profiles. With the aim to
understand the cosmic acceleration or dark energy phenomena, many
theoretical models have been proposed, like the scalar field models
and the modified Einstein gravity models \cite{r1} \cite{r2}
\cite{r3}.

Due to the limited scope of our experimental and observational
tools, we have not yet been able to understand  the ``dark'' nature
and to detect the origin of DM and DE. Purely gravitational probes
can not provide enough information to differentiate these two kinds
of mysterious constitutions either. Therefore, from the
phenomenological and practical point of view, a single (unified DM
with DE) fluid description may be more plausible at least in the
cosmic evolution description, which utilizes a single equation of
state to model the dark matter and dark energy contributions
together \cite{df1} \cite{df2} \cite{df3} \cite{df4} \cite{df5}
\cite{Liddle}. Generally, such models has a non-constant equation of
state (EoS), which reflects both its dynamical and thermodynamics
characters. The density dependent equation of state is widely
investigated, such as the famous Chaplygin gas and generalized
Chaplygin gas models, which assume an EoS form like the
$p=-A/\rho^{\alpha}$ where the $\alpha$ is a model parameter.
Another practical method to modify the EoS is by the introduction of
cosmic viscosity media contribution which replaces the simplest
perfect fluid EoS on a more physical and realistic basis. In the
homogeneous and isotropic Friedmann-Robertson-Walker frame, only a
bulk viscosity term which behaves as an additive pressure
contribution can mimic both the two dark components and their
coupling effects by playing a main role to influence the cosmic
evolution. Different forms of the viscosity coefficients have been
proposed like the density $\rho$ dependent \cite{d} or the redshift
$z$ dependent \cite{md}.

In this letter, we will proceed our effort on the investigation of
the viscosity dark fluid model to study the scalar perturbation of
this model, for perturbation analysis can provide us a powerful tool
to differentiate and constrain cosmology models finely as the
calculation of perturbation quantities links the theoretical models
with more plentiful and precise observations, like the cosmic
microwave background (CMB) and large scale structure observation.
There have been some researches on the perturbation evolution of the
viscosity models \cite{FGR} \cite{lb} \cite{z}, by which we know
that after corresponding model parameters chosen properly, the
Chaplygin gas formulation can be viewed as a special case of the
density dependent viscosity model. In the non-perturbative (zero
order) level, the Chaplygin gas model can be exactly solved and fit
the observational data well. But it has been found that in the
perturbation level, there exist some unacceptable behaviors, like
the blow up of density perturbation evolution and other peculiar
behaviors \cite{z}  \cite{ps}. One motivation to build other kinds
of the viscosity models is to overcome these difficulties the
Chaplygin gas models possess. Here, we will consider a
time-dependent viscosity coefficient model, which is equivalent to
the introduction of a general Equation of state(EoS) \cite{RM}. The
general EoS is
\begin{equation}\label{1}
p=(\gamma-1)\rho+p_{0}+w_{H}H+w_{H2}H^{2}+w_{dH}\dot H.
\end{equation}
In the background, this model can fit the current astrophysics
observational datasets consistently. We derive its perturbation
equations that govern the evolution of gravitational potential and
density perturbation below. We numerically solve the perturbation
equation, and compare it with that of conventional $\Lambda$CDM
model and the Chaplygin gas model finding that the dark fluid model
behaves well in different scales. Though there exists some value
difference between the $\Lambda$CDM model and the dark fluid model
in the late time evolution, their gravitational potential and
density contrast shape and evolution behavior are similar by
plotting respectively.

This paper is organized as follows: In Sec. \textbf{II}, we
summarize the calculations of scalar perturbation, and give the
general evolution equation of the gravitational potential. In Sec.
\textbf{III}, we briefly review the background evolution of the dark
fluid model. In Sec. \textbf{IV}, we discuss the perturbation
evolution of the dark fluid model. In Sec. \textbf{V}, we
numerically solve the perturbation equation and compare it with
other models. Finally, we present the conclusions in the last
section.
\section{Calculations of scalar perturbation}
In this paper, we choose Newtonian gauge to calculate the scalar
perturbation
\begin{equation}
\rd s^{2}=-(1+2\phi)\rd t^{2}+a(t)^{2}\delta_{ij}(1-2\psi)\rd
x^{i}\rd x^{j}.
\end{equation}
If making the assumption here that there is no contribution from
anisotropy inertia, it concludes that $\phi=\psi$.

Generally, Einstein field equation with perturbed metric takes the
form(for simplicity, we set $\kappa=1$ hereafter.) \cite{c1}
\cite{c2}
\begin{equation}\label{2} \frac{\dot
a}{a}\dot\phi+\frac{\dot
a^{2}}{a^{2}}\phi-\frac{1}{3}\frac{\nabla^{2}}{a^{2}}\phi=-\frac{1}{6}\delta\rho,
\end{equation}
\begin{equation}\label{3}
\dot\phi+\frac{\dot a}{a}\phi=-\frac{1}{2}(\rho+p)\delta u,
\end{equation}
\begin{equation}\label{4}
\ddot\phi+3\frac{\dot a}{a}\dot\phi+(2\frac{\ddot
a}{a}+\frac{1}{3}\frac{\nabla^{2}}{a^{2}})\phi=\frac{1}{6}(\delta\rho+3\delta
p),
\end{equation}
where $\delta\rho$ and $\delta p$ are first order perturbation to
zero-order cosmic density $\rho$ and pressure $p$ respectively.
Perturbation to velocity of cosmic fluid $\delta u_{i}$ is
decomposed as $\delta u_{i}=\nabla_{i}\delta u+\delta u'_{i}$.
$\delta u$ is the scalar velocity potential. $\delta u'_{i}$ is a
divergenceless vector, which we also assume here contributes no
effect. $\nabla$ denotes gradient with respect to comoving
coordinate. From Eq. (\ref{2}) and (\ref{3}), we obtain a constrain
on first-order perturbation quantity $\phi$, $\delta\rho$ and
$\delta u$
\begin{equation}
2\frac{\nabla^{2}}{a^{2}}\phi-\delta\rho+3\frac{\dot
a}{a}(\rho+p)\delta u=0.
\end{equation}
Also momentum and energy conservation equation to first order in
perturbation could be derived
\begin{equation}
\delta p+\partial_{t}[(\rho+p)\delta u]+\frac{\dot
a}{a}(\rho+p)\delta u+(\rho+p)\phi=0,
\end{equation}
\begin{equation}
\dot\delta\rho+\frac{3\dot a}{a}(\delta\rho+\delta
p)+\nabla^{2}\big[\frac{1}{a^{2}}(\rho+p)\delta
u\big]-3(\rho+p)\dot\phi=0.
\end{equation}
A generalized parameterized equation of state may have the form as
\begin{equation}
p=(\gamma-1)\rho+f(\rho;\dot\rho;\alpha_{i})+...
\end{equation}
or
\begin{equation}
p=(\gamma-1)\rho+g(H;\dot H;\beta_{i})+...,
\end{equation}
where $\alpha_{i}$ and $\beta_{i}$ are model parameters. When the
models are discussed in flat universe $k=0$, two parameterized
function $f(\rho)$ and $g(H)$ are categorized in one class. The
equation of state reduces to the perfect fluid case with EoS as
$p=(\gamma-1)\rho$ when the model parameters $\alpha_{i}$ vanish.

For a barotropic equation of state, adiabatic sound speed is defined
as $c^{2}_{a}=\frac{\rd p}{\rd\rho}$, hence the ratio of pressure
and density perturbation is $c^{2}_{a}$. Therefore we could
eliminate $\delta\rho$ from Eq. (\ref{2}) and (\ref{4}) ,then obtain
the equation governs the gravitational potential
\begin{equation}\label{ge}
\ddot\phi+(4+3c^{2}_{a})\frac{\dot a}{a}\dot\phi+[2\frac{\ddot
a}{a}+(1+3c^{2}_{a})(\frac{\dot
a}{a})^{2}-c^{2}_{a}\frac{\nabla^{2}}{a^{2}}]\phi=0
\end{equation}
If write metric perturbation as Fourier integral
\begin{equation}
\phi(\textbf{x},t)=\int\rd^{3}q\,\mathrm{e}^{i\textbf{q}\cdot\textbf{x}}\phi_{q}(t),
\end{equation}
we can make the substitution $\nabla^{2}\rightarrow q^{2}$, where
$q$ is the wave number.

\section{Background evolution}

In this section, we briefly review the background evolution behavior
of the viscosity dark fluid model. A general form of EoS
investigated in \cite{RM} is
\begin{equation}
p=(\gamma-1)\rho+p_{0}+w_{H}H+w_{H2}H^{2}+w_{dH}\dot H
\end{equation}
One can prove that this generally parameterized EoS can be
effectively equivalent to a single fluid with a time-dependent bulk
viscosity:
\begin{equation}
\zeta=\zeta_{0}+\zeta_{1}\frac{\dot a}{a}+\zeta_{2}\frac{\ddot
a}{\dot a},
\end{equation}
and three parameters in the viscosity coefficient correspond to EoS
parameters as
\begin{subequations}
\begin{eqnarray}
w_{H} &=& -3\zeta_{0},\\
w_{H_{2}} &=& -3(\zeta_{1}+\zeta_{2}),\\
w_{dH} &=& -3\zeta_{2}.
\end{eqnarray}
\end{subequations}
Flat Friedmann-Robertson-Walker metric reads
\begin{equation}
\rd s^{2}=-\rd t^{2}+a(t)^{2}\delta_{ij}\rd x^{i}\rd x^{j}.
\end{equation}
The energy-momentum tensor with modified EoS could be written as
\begin{equation}
T_{\mu\nu}=\rho U_{\mu}U_{\nu}+\tilde pH_{\mu\nu},
\end{equation}
where $\tilde p$ represents modified pressure, in model concerned it
is Eq.(\ref{1}), and in comoving coordinate $U^{\mu}=(1,0,0,0)$ .
Due to the correspondence between this modified EoS model and
viscosity model, pressure could also be $\tilde p=p-\zeta\theta$,
where $\theta=U^{\mu}_{;\mu}=3\dot a/a$.

Using the EoS above and Friedmann equation, the equation of scale
factor $a(t)$ evolution could be obtained
\begin{equation}
\frac{\ddot
a}{a}=\frac{-(3\gamma-2)/2-(\kappa^{2}/2)w_{H2}+(\kappa^{2})w_{dH}}{1+(\kappa^{2})w_{dH}}\big(\frac{\dot
a}{a}\big)^{2}+\frac{-(\kappa^{2})w_{H}}{1+(\kappa^{2}/2)w_{dH}}\frac{\dot
a}{a}+\frac{-(\kappa^{2}/2)p_{0}}{1+(\kappa^{2}/2)w_{dH}}.
\end{equation}
After redefining model parameters, there will be a compact form of
evolution equation, at the same time, this form is comparable to
perfect fluid case
\begin{equation}\label{sf}
\frac{\ddot a}{a}=-\frac{3\tilde\gamma-2}{2}\big(\frac{\dot
a}{a}\big)^{2}+\frac{1}{T_{1}}\frac{\dot a}{a}+\frac{1}{T^{2}_{2}},
\end{equation}
where
\begin{equation}\label{9}
\tilde\gamma=\frac{\gamma+(\kappa^{2}/3)w_{H2}}{1+(\kappa^{2}/2)w_{dH}},
\end{equation}
\begin{equation}
\frac{1}{T_{1}}=\frac{-(\kappa^{2}/2)w_{H})}{1+(\kappa^{2}/2)w_{dH}},
\end{equation}
\begin{equation}
\frac{1}{T^{2}_{2}}=\frac{-(\kappa^{2}/2)p_{0}}{1+(\kappa^{2}/2)w_{dH}},
\end{equation}
\begin{equation}\label{12}
\frac{1}{T^{2}}=\frac{1}{T^{2}_{1}}+\frac{6\tilde\gamma}{T^{2}_{2}}.
\end{equation}
The solution of scale factor is
\begin{eqnarray}
a(t) & = &
a_{0}\big\{\frac{1}{2}\big(1+\tilde{\gamma}\theta_{0}T-\frac{T}{T_{1}}\big)\mathrm{exp}\big[\frac{t-t_{0}}{2}\big(\frac{1}{T}+\frac{1}{T_{1}}\big)\big]+{}
                                                                                                                        \nonumber\\ & &
{}\frac{1}{2}\big(1-\tilde{\gamma}\theta_{0}T+\frac{T}{T_{1}}\big)\mathrm{exp}\big[-\frac{t-t_{0}}{2}\big(\frac{1}{T}-\frac{1}{T_{1}}\big)\big]\big\}^{2/3\tilde{\gamma}}.
\end{eqnarray}
From Friedmann equation, cosmic density evolution reads
\begin{equation}
\rho(t)=\frac{1}{3\kappa^{2}\tilde{\gamma}^{2}}\bigg[\frac{(1+\tilde{\gamma}\theta_{0}T-\frac{T}{T_{1}})(\frac{1}{T}+\frac{1}{T_{1}})\mathrm{exp}(\frac{t-t_{0}}{T})-(1-\tilde{\gamma}\theta_{0}T+\frac{T}{T_{1}})(\frac{1}{T}-\frac{1}{T_{1}})}{(1+\tilde{\gamma}\theta_{0}T-\frac{T}{T_{1}})\mathrm{exp}(\frac{t-t_{0}}{T})+(1-\tilde{\gamma}\theta_{0}T+\frac{T}{T_{1}})}\bigg]^{2}.
\end{equation}
The case above is for $\tilde{\gamma}\neq0$, when take the limit of
$\tilde{\gamma}$, solution could be obtained
\begin{equation}
a(t)=a_{0}\mathrm{exp}\big[\big(\frac{1}{3}\theta_{0}T_{1}+\frac{T^{2}_{1}}{T^{2}_{2}}\big)\big(\mathrm{exp}(\frac{t-t_{0}}{T_{1}})-1\big)-\frac{T_{1}(t-t_{0})}{T^{2}_{2}}\big],
\end{equation}
and cosmic density evolution
\begin{equation}
\rho(t)=\frac{3}{\kappa^{2}}\big[\frac{1}{3}\theta_{0}\mathrm{exp}(\frac{t-t_{0}}{T_{1}})+\frac{T_{1}}{T^{2}_{2}}(\mathrm{exp}(\frac{t-t_{0}}{T_{1}})-1)\big].
\end{equation}

Using Friedmann equation, Eq. (\ref{1}) could be converted into a
form, r.h.s of which is only the function of density $\rho$
\begin{equation}\label{eos}
p=(\tilde{\gamma}-1)\rho-\frac{2}{T_{1}}\sqrt{\frac{\rho}{3}}-\frac{2}{T^{2}_{2}},
\end{equation}
where parameters are defined the same as (\ref{9})-(\ref{12}).

\section{Evolution of Scalar perturbation}
\subsection{Evolution equation}
In adiabatic perturbation case, ratio of adiabatic density and
pressure perturbation equals to the adiabatic sound speed
\begin{equation}
\frac{\delta p_{a}}{\delta\rho_{a}}=c^{2}_{a}.
\end{equation}
In this paper we pay our attention on single fluid model and
investigate perturbation in adiabatic region, that is, pressure
perturbation is proportional to the density perturbation. For a
barotropic EoS model, the proportional efficient between pressure
and density perturbation is merely a function of density $\rho$. On
the other hand, in the condition of single (dark) fluid, we assume
that dark sector interacts with baryon matter and the least
dominated radiation negligibly.

With the barotropic EoS (\ref{eos}), adiabatic sound speed is
$c^{2}_{a}=\tilde{\gamma}-1-\frac{1}{\sqrt{3}T_{1}}\frac{1}{\sqrt{\rho}}$.
From Eq. (\ref{ge}), we obtain the correspondent equation of
single(dark) fluid model
\begin{equation}\label{25}
\ddot\phi_{q}+f\dot\phi_{q}+g\phi_{q}=0,
\end{equation}
where two coefficients are defined by
\begin{equation}
f(H)=(3\tilde{\gamma}+1)H-\frac{1}{T_{1}}
\end{equation}
and
\begin{equation}
g(H;q)=2\dot
H+3\tilde{\gamma}H^{2}-\frac{1}{T_{1}}H-(\tilde{\gamma}-1-\frac{1}{3T_{1}H})\frac{q^{2}}{a^{2}}.
\end{equation}
Conveniently, one can decompose metric perturbation as
$\phi_{q}(t)=v(t)p(t)$, therefore obtains a differential equation
about  $v(t)$ and $p(t)$
\begin{equation}\label{28}
\ddot v+(2\frac{\dot p}{p}+f)\dot v+(\frac{\ddot p+f\dot
p+gp}{p})v=0.
\end{equation}
If we choose the function $p(t)$ properly as
\begin{equation}\label{p}
p(t)=\mathrm{exp}\big(-\frac{1}{2}\int^{t}_{0}\rd t'f(t')\big),
\end{equation}
then the damping term can be eliminated, so Eq. (\ref{28}) reduces
to
\begin{equation}\label{30}
\ddot v-(\frac{1}{2}\dot f+\frac{1}{4}f^{2}-g)v=0.
\end{equation}
which takes a harmonic oscillator form. For the short wave limit, we
have $\ddot v+g(q;H)v=0$ and
$g\approx-(\tilde{\gamma}-1-\frac{1}{3T_{1}H})\frac{q^{2}}{a^{2}}$.
If WKB condition is fulfilled, we have the approximate solution
\begin{equation}
v\approx\frac{1}{g(q;H)^{1/4}}\big\{c_{+}\mathrm{exp}\big(i\int\rd
t\sqrt{g(q;H)}\big)+c_{-}\mathrm{exp}\big(-i\int\rd
t\sqrt{g(q;H)}\big)\big\}
\end{equation}
\subsection{The future solution of the gravitational potential}
Eq. (\ref{25}) is too complicated to solve exactly, so here we will
consider a simpler asymptotic case and try to extract the solution
in this limit. If we strict that parameter $T_{1}$ should be
negative to confirm that sound speed is real, then we could see from
Eq. (19) in {\cite{RM}} that cosmic density approaches a constant
value
\begin{equation}
\rho=-\frac{3T_{1}}{T^{2}_{2}}
\end{equation}
as $t\rightarrow\infty$, and negative $T_{1}$ also confirms the
positive of energy. It concludes that the universe will enter a de
Sitter period then: $H\rightarrow H_{\Lambda}$, where $H_{\Lambda}$
is a constant. Also the scale factor evolves exponentially:
\begin{equation}\label{32}
a(t)=\mathrm{e}^{H_{\Lambda}t} \qquad
H_{\Lambda}=\sqrt{-\frac{T_{1}}{T^{2}_{2}}},
\end{equation}
Eq. (\ref{30}) becomes
\begin{equation}
\ddot v-[\frac{1}{4}f^{2}_{\Lambda}-g_{\Lambda}(q)]v=0,
\end{equation}
where $f_{\Lambda}=(3\tilde{\gamma}+1)H_{\Lambda}-\frac{1}{T_{1}}$
is the late time asymptotic constant of function $f(H)$ and
$g_{\Lambda}(q)=3\tilde{\gamma}H_{\Lambda}^{2}-\frac{1}{T_{1}}H_{\Lambda}-(\tilde{\gamma}-1-\frac{1}{3T_{1}H_{\Lambda}})\frac{q^{2}}{3a^{2}}$
is the function of $g(H;q)$ when Hubble parameter approaches
constant. Then we can solve the differential equation (\ref{32}),
and get
\begin{equation}
v(t)=
C_{1}J_{m}\big(\frac{\sqrt{na^{-1}(t)}}{H_{\Lambda}}\big)\Gamma\big(1-\frac{i\sqrt{m}}{H_{\Lambda}}\big)+C_{2}J_{-m}\big(\frac{\sqrt{na^{-1}(t)}}{H_{\Lambda}}\big)\Gamma\big(1+\frac{i\sqrt{m}}{H_{\Lambda}}\big),
\end{equation}
where two parameter $m$ and $n$ are defined for simplicity by
\begin{equation}
m=\frac{1}{4}f^{2}_{\Lambda}-3\tilde{\gamma}H_{\Lambda}^{2}+\frac{1}{T_{1}}H_{\Lambda}=\frac{1}{2}\big[(3\tilde{\gamma}-1)^{2}H_{\Lambda}^{2}-\frac{1}{T_{1}}]^{2},
\end{equation}
\begin{equation}
n=-\frac{1}{3}(\tilde{\gamma}-1-\frac{1}{3T_{1}H_{\Lambda}})q^{2}.
\end{equation}

When the physical wavelength is much longer than the Hubble radius
\begin{equation}
\lambda_{phys}\gg H^{-1}\Rightarrow\frac{q}{a(t)}\ll H.
\end{equation}
The solution in this long wave limit could be obtained directly from
Eq. (\ref{25}). Assuming the solution takes the form as
\begin{equation}
\phi_{q}(t)=\mathrm{e}^{bt}.
\end{equation}
After inserting it into Eq. (\ref{25}), we get
\begin{equation}
b^{2}+[(3\tilde{\gamma}+1)H_{\Lambda}-\frac{1}{T_{1}}]b+3\tilde{\gamma}H_{\Lambda}^{2}-\frac{1}{T_{1}}H_{\Lambda}=0.
\end{equation}
This quadratic equation has two dependent solution, therefore the
solution of Eq.(\ref{25}) reads
\begin{equation}
\phi_{q}=c_{1}\mathrm{exp}\big[(-H_{\Lambda}+\frac{1-H_{\Lambda}}{T_{1}})t\big]+c_{2}\mathrm{exp}\big[(-6\tilde{\gamma}H_{\Lambda}+\frac{1+H_{\Lambda}}{T_{1}})t\big]
\end{equation}
This unstable solution exponentially increases or decreases in the infinite future,
which dependents on the parameters.

For the opposite limit, we consider large wave number solution with $q\gg aH$, and fix at some
conformal wave number $k=q/a$. Then we have the simple solution in
this limit
\begin{equation}
v\propto e^{\sqrt{lk}t},
\end{equation}
where $l=\tilde\gamma-1-\frac{1}{3T_{1}H}$. Together with the
definition (\ref{p}), we see this is an exponential decay solution
for the gravitational potential, which is also unstable.
\section{Numerical results}
\subsection{Comparison models}
In the numerical results presented below, we also calculate
perturbation in $\Lambda$CDM model and the Chaplygin gas model
numerically as a
comparison, so we present a short review here.\\
(a) $\Lambda$CDM model: The cosmological constant does not
contribute perturbation in the total energy density
\begin{equation}
\delta\rho_{\Lambda}=0.
\end{equation}
The density perturbation comes from matter density,
$\delta\rho=\delta\rho_{m}$. The cosmological constant plays a role
in influencing the background evolution of the universe, especially
by Hubble parameter, so it will be imprinted into the evolution of
matter perturbation. We could obtain the scale-independent gravity
perturbation equation from Eqs. (\ref{2}) and (\ref{4}):
\begin{equation}\label{l}
\ddot\phi_{mq}+4H\dot\phi_{mq}+(2\dot H+3H^{2})\phi_{mq}=0.
\end{equation}
In the matter dominated era
\begin{equation}
H^{2}=\frac{1}{3}\rho_{m}=\Omega_{m0}a^{-3}=\frac{2}{3}t^{-2}.
\end{equation}
Also the solution $\phi_{mq}=\phi_{mq0}+\frac{3c}{5}t^{-5/3}$ is
well-known.\\
(b) Chaplygin gas model: It has the EoS form
\begin{equation}
p=-\frac{A}{\rho}
\end{equation}
If we use this EoS to model single fluid universe, then get Hubble
parameter
\begin{equation}
H(z)=[\tilde\Omega(1+z)^{6}+(1-\tilde\Omega)]^{1/4}
\end{equation}
and the adiabatic sound speed
\begin{equation}
c^{2}_{a}=\frac{A}{\rho^{2}}
\end{equation}
Hence $H$ and $c^{2}_{a}$ in Eq. (\ref{25}) are specified.
\subsection{Effective dark fluid}
\subsubsection{Background}
For simplicity, we set parameter $T_{2}\rightarrow\infty$, which we
call this case effective dark fluid. This means if
$\rho\rightarrow0$, pressure $p$ vanishes too, and there is not a
cosmological constant like pressure contribution, which could be
seen from EoS (\ref{eos}). Hence Eq. (\ref{sf}) becomes
\begin{equation}\label{t20}
\frac{\ddot a}{a}=-\frac{3\tilde\gamma-2}{2}\big(\frac{\dot
a}{a}\big)^{2}+\frac{1}{T_{1}}\frac{\dot a}{a},
\end{equation}
This differential equation about scale factor can be seen as a
special case given by a general EoS \cite{eos1} \cite{eos2}
\cite{eos3}
\begin{equation}
p=-\rho-A\rho^{\alpha}-BH^{2\beta},
\end{equation}
which can give
\begin{equation}
\frac{\ddot a}{a}=-\frac{3\tilde\gamma-2}{2}\big(\frac{\dot
a}{a}\big)+\lambda\big(\frac{\dot a}{a}\big)^{m}+\mu\big(\frac{\dot
a}{a}\big)^{n}+\nu.
\end{equation}
Eq. (\ref{t20}) can be converted as a differential equation of
$H(a)$
\begin{equation}
a\frac{\rd H}{\rd a}=-\frac{3\tilde\gamma}{2}H+\frac{1}{T_{1}}.
\end{equation}
Its solution is
\begin{equation}
H(a)/H_{0}=\Omega a^{-3\tilde\gamma/2}+(1-\Omega),
\end{equation}
where $\Omega=1-\frac{2}{3\tilde\gamma T_{1}H_{0}}$, and we have
already set $a_{0}=1$. We use this Hubble parameter with
$\tilde\gamma=0.9$ and $\tilde\gamma=1.2$ to calculate the distance
\begin{equation}
D_{L}(z)=H_{0}(1+z)\int^{z}_{0}\frac{1}{H(z')}\rd z'
\end{equation}
and the distance modulus. We compare it with the supernova data
\cite{sd}, which is plotted in FIG. 1. We see that model with
parameter $\tilde\gamma=0.9$ (blue) and $\tilde\gamma=1.2$ nearly
can not be discriminated in the late time (small redshift, more data
in this region has been obtained). There are some differences for
larger redshift. But in the whole, both two parameters consist with
the data well.
\begin{figure}[htp]
\centering
\includegraphics[scale=0.8]{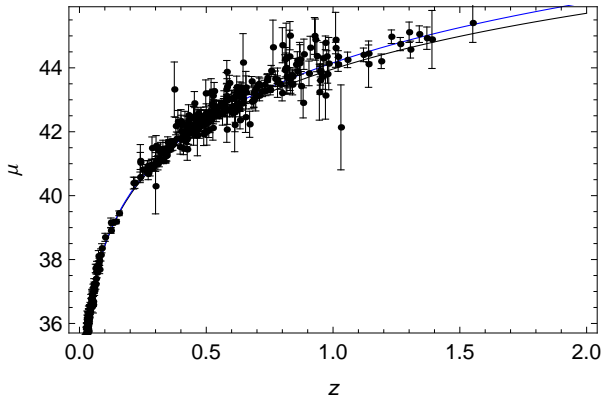}
\caption{The relation between distance modulus and redshift. Black
and Blue lines correspond to the theoretical calculation curves with
$\tilde\gamma=1.2$ and $0.9$ respectively. Both of them can fit data
in an acceptable level.}
\end{figure}
\subsubsection{The gravitational potential and the density perturbation}
After inserting the Hubble parameter into Eq. (\ref{25}), we
numerically solve this scale-dependent model. Results are
illustrated in FIG. 2. As a comparison, $\Lambda$CDM and Chaplygin
gas model are solved and plotted (green and red line respective)
too. In the early time, the difference between $\Lambda$CDM and dark
fluid model is tiny. Both of them behaves nearly as a constant. In
the late time, two models give the same shape of the gravitational
potential. Though these two models predicts the decay of potential,
there exists a value contrast around $5\%-10\%$. The quantity of
contrast dependents on the value of model parameter $\tilde\gamma$.
Here we point that the dark fluid model predicts more similar
potential as $\Lambda$CDM than the Chaplygin gas model. On the other
hand, it can be qualitatively seen the level of scale-dependent of
the dark fluid model. We plot different results with $q=0.005$,
$0.5$ and $1.5$.

For the small scale, the perturbation evolution of the dark fluid
model with $\tilde\gamma=0.9$ is significantly different. The
gravitational potential decays much earlier, which contradicts with
$\Lambda$CDM. Numerical results for $\tilde\gamma<1$ indicate that
this condition strongly influences the early evolution of
perturbation quantity(we only plot $\tilde\gamma=0.9$ case here.),
which puts strict constrain on parameter region. Also, if
$\tilde\gamma$ is bigger than $1$, the positive of adiabatic sound
speed could be easily fulfilled.
\begin{figure}
\centering
\includegraphics[scale=0.5]{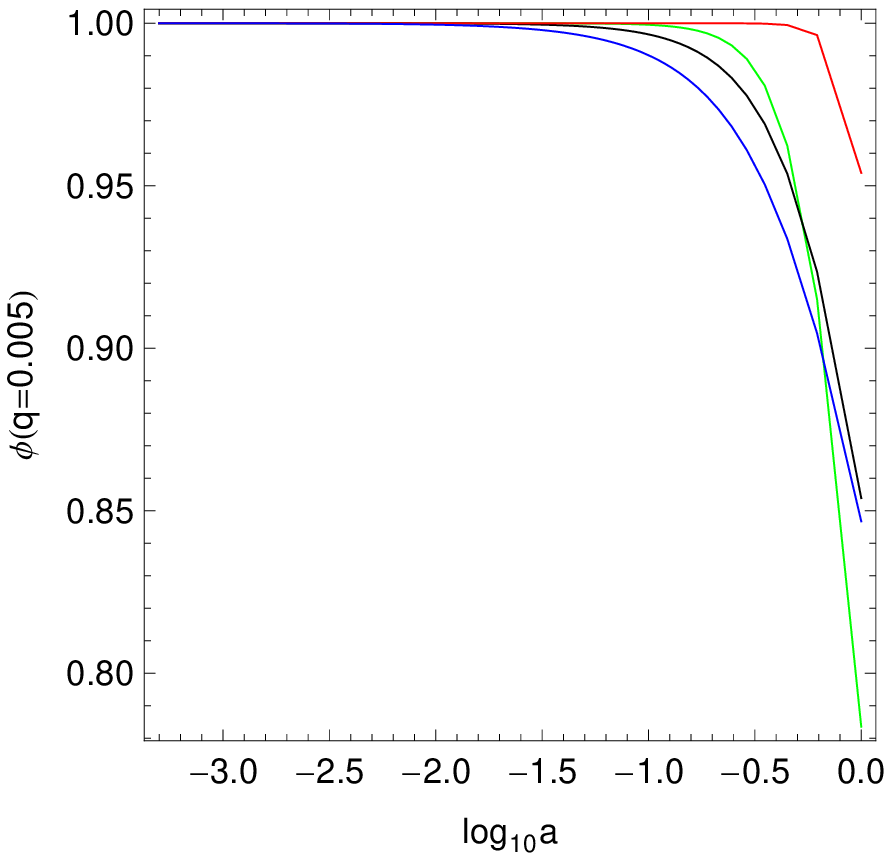}
\includegraphics[scale=0.5]{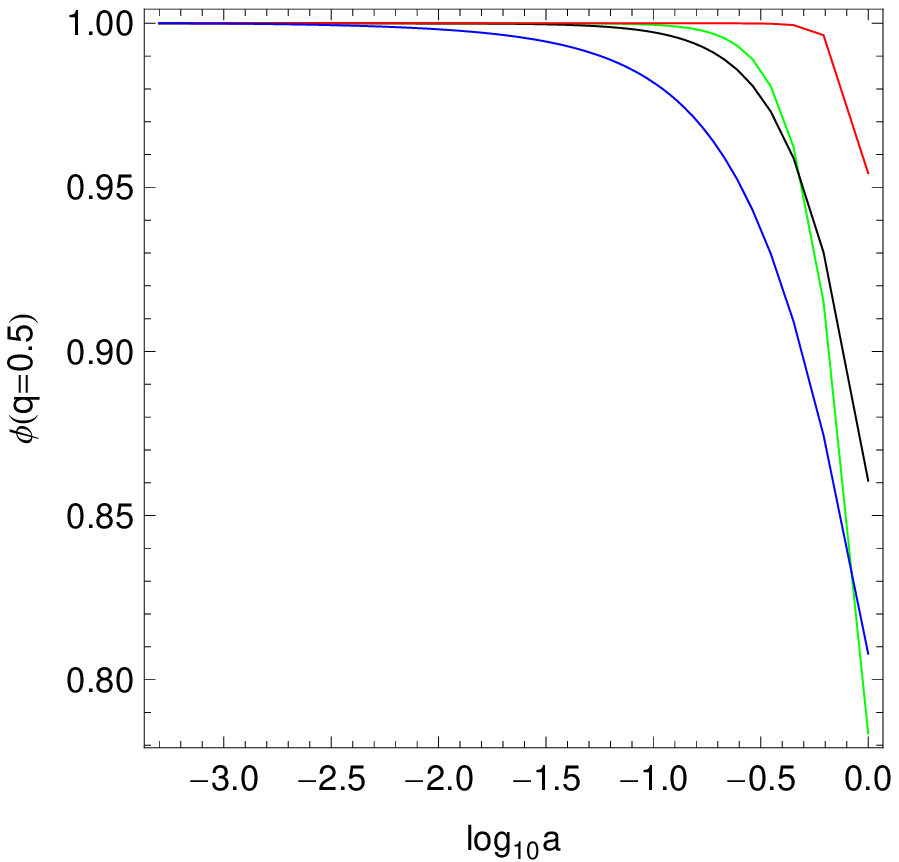}
\includegraphics[scale=0.5]{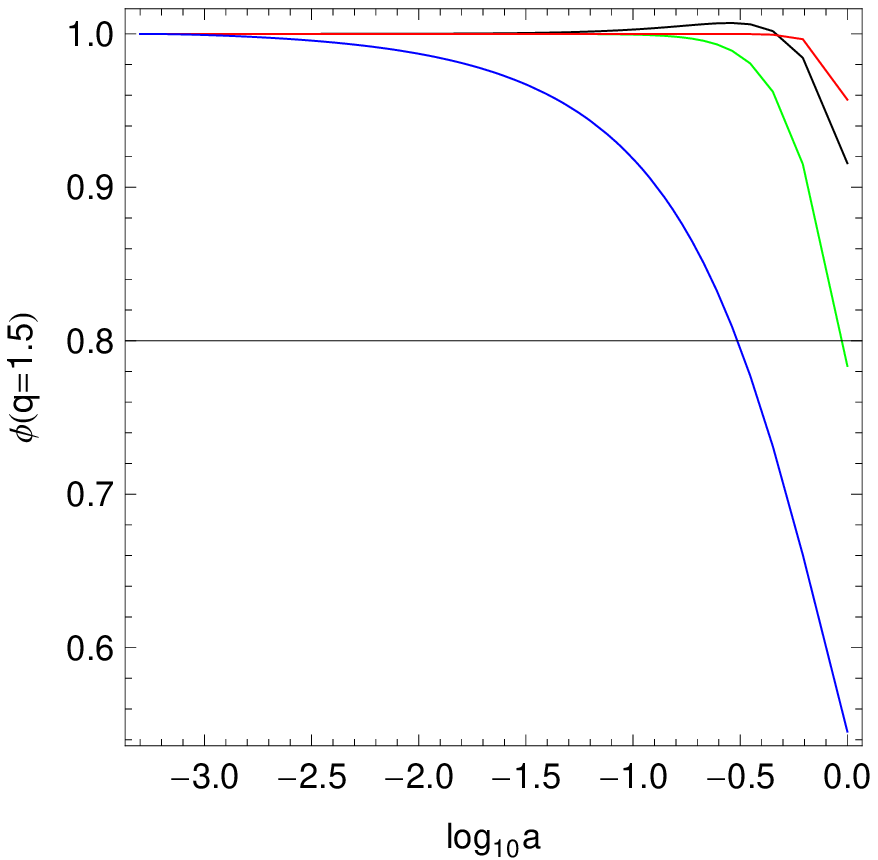}
\caption{The gravitational potential for different wave
number($q=0.005, 0.5, 1.5$) evolves with time. Black, blue, green
and red lines represent the dark fluid model with $\tilde\gamma=1.2$
and $0.9$, $\Lambda$CDM and the Chaplygin gas model respectively.}
\end{figure}

From Eq. (\ref{2}), density perturbation could be expressed as
\begin{equation}
\delta\rho=-6H\dot\phi-6H^{2}\phi+2\frac{q^{2}}{a^{2}}\phi.
\end{equation}
It tells us that once we have the solution of the gravitational
potential we could get the information of the density contrast.
During the early time ($\dot\phi\simeq0$) and in the very large
scale ($q\ll aH$), the density contrast and the gravitational
potential is linked by the Hubble parameter
$\delta\rho\simeq-6H^{2}\phi$. Always define
\begin{equation}
\delta=\frac{\delta\rho}{\rho}=-2\frac{\dot{\phi}}{H}-2\phi+\frac{2}{3}\frac{q^{2}}{a^{2}H^{2}}\phi,
\end{equation}
where Friedmann equation is used in the last step.
\begin{figure}[htp]
\centering
\includegraphics[scale=0.5]{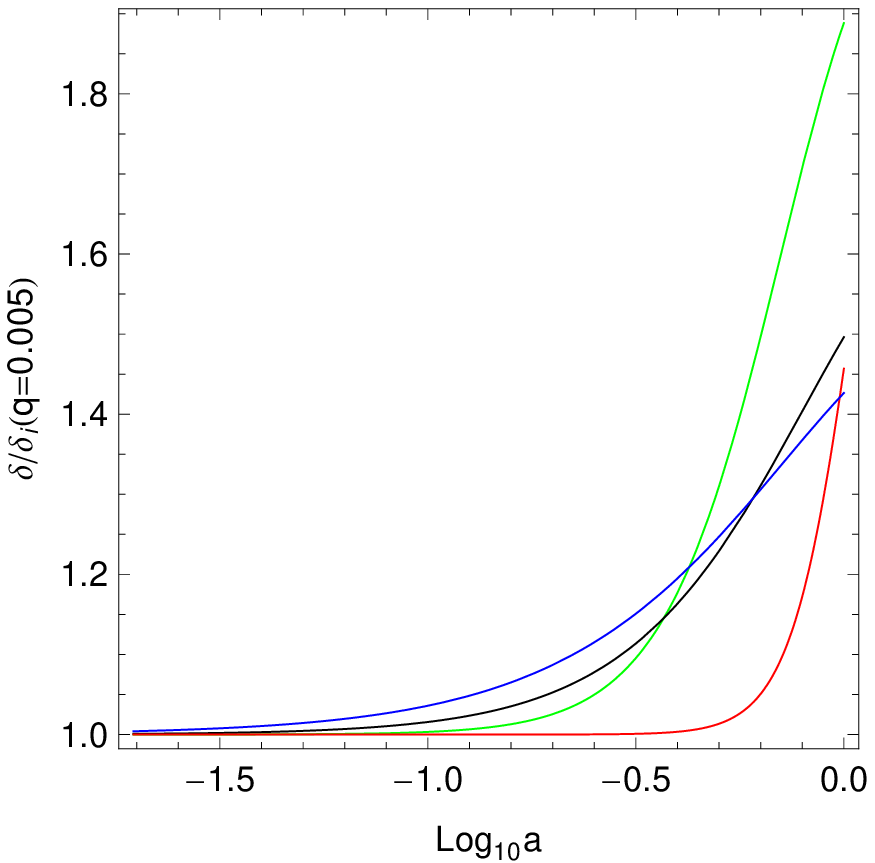}
\includegraphics[scale=0.5]{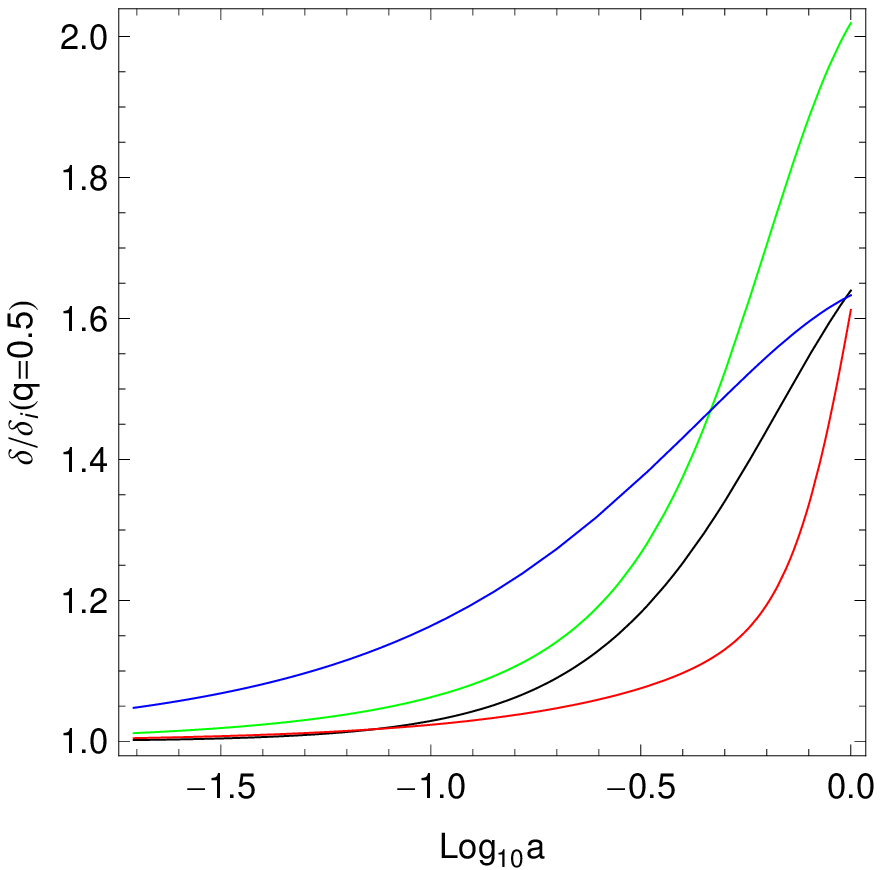}
\includegraphics[scale=0.5]{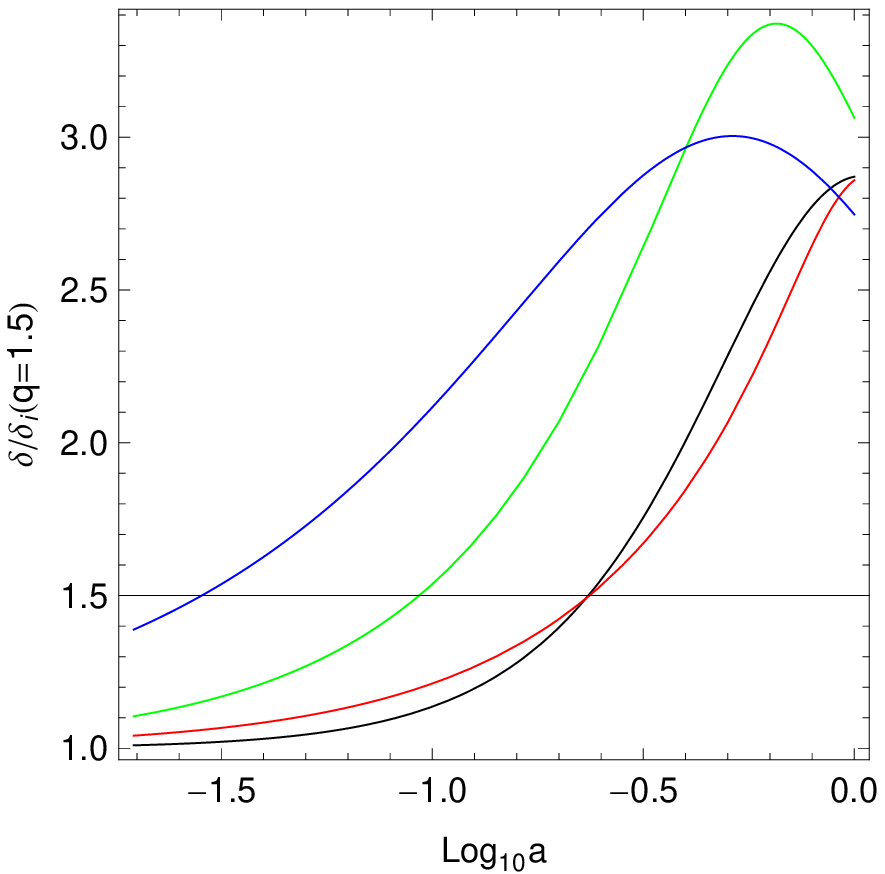}
\caption{Density perturbation with $q=0.005, 0.5$ and $1.5$ evolves
with time. Black, blue, green and red lines represent the dark fluid
model with $\tilde\gamma=1.2$ and $0.9$, $\Lambda$CDM and the
Chaplygin gas model respectively.}
\end{figure}
Numerical results of the density perturbation $\delta$ for different
$q$ is plotted in FIG. 3. The numerical curves have the similar
shape, but for modified models, the density perturbation is
suppressed in the late time. For the large scale, the density
perturbation evolution in the dark fluid model increases linearly in
the late universe. $\delta$ deviates from that of $\Lambda$CDM in
the late time, and suppressed today for different scale. The values
enhanced or depressed dependent on parameter $\tilde\gamma$ and
scale.

\section{Conclusion and Discussion}
In this paper, we investigate extensively the dark fluid model
proposed in \cite{RM}, equivalently this model can be viewed as a
single fluid with time-dependent bulk viscosity. Scale factor and
density evolution can be exactly solved in this model. In the
background, the dark fluid model can fit the supernova data
acceptable. Our main task in this paper is to analysis the behavior
of this model in the perturbation level. We derive equations govern
the perturbation quantities. For the condition that $T_{1}$ is
smaller than $0$, the universe will enter de Sitter phase in the
$t\rightarrow\infty$ future. We solve exactly the gravitational
equation in this condition and obtain the solution for both long and
short wave case. Generally, the perturbation evolution equations are
solved numerically. When compare the results with those of
$\Lambda$CDM model, we find that
\begin{itemize}
\item In the early time and the large scale, both the gravitational potentials of two models
behave as a constant.
\item Though the gravitational potentials of two models have similar
behavior and shape, as can be seen form FIG. ~3, there exists about
$5\%-10\%$ significant value difference in the late time.
\end{itemize}

Perturbation analysis also provides constraint on model parameter.
For different selection of parameter $\tilde\gamma$, both
$\tilde\gamma<0$ and $\tilde\gamma>0$ can give consistent prediction
curve of distance modulus, but numerical results indicate in
$\tilde\gamma<0$ case, the gravitational potential deviates from
$\Lambda$CDM significantly from the early time. This result strongly
constrain the selection region of $\tilde\gamma$. We suggest
$\tilde\gamma>0$, which can also produce positive sound speed
naturally.

\section*{Acknowledgements}

This work is partly supported by NSF of China under Grant
No.10675062 and by the project of knowledge Innovation Program
(PKIP) of Chinese Academy of Sciences under Grant No.KJCX2.YW.W10
through the KITPC.

\end{document}